\documentclass[aps,preprint]{revtex4}%
\usepackage{amsfonts}
\usepackage{amsmath}
\usepackage{amssymb}
\usepackage{graphicx}%
\providecommand{\U}[1]{\protect\rule{.1in}{.1in}}

\begin{document}
\title{Dirac's Classical-Quantum Analogy for the Harmonic Oscillator: Classical
Aspects in Thermal Radiation Including Zero-Point Radiation}
\author{Timothy H. Boyer}
\affiliation{Department of physics, City College of the City University of New York, New
York, New York 10031}

\begin{abstract}
Dirac's Poisson-bracket-to-commutator analogy for the transition from
classical to quantum mechanics assures that for many systems, the classical
and quantum systems share the same algebraic structure. \ The quantum side of
the analogy (involving operators on Hilbert space with commutators scaled by
Planck's constant $\hbar)$ not only gives the algebraic structure but also
dictates the average values of physical quantities in the quantum ground
state. \ On the other hand, the Poisson brackets of nonrelativistic mechanics,
which give only the classical canonical transformations, do not give any
values for physical quantities. \ Rather, one must go outside nonrelativistic
classical mechanics in order to obtain a fundamental phase space distribution
for classical physics. \ We assume that the values of physical quantities in
classical theory at any temperature depend on the phase space probability
distribution which arises from thermal \textit{radiation}
equilibrium\ including classical zero-point radiation with the scale set by
Planck's constant $\hbar.~\ $All mechanical systems in thermal radiation will
inherit the constant $\hbar$ from thermal radiation. \ Here we note the
connections between classical and quantum theories \ (agreement and contrasts)
at all temperatures for the harmonic oscillator in one and three spatial
dimensions. \ \ 

\end{abstract}
\maketitle

\section{Classical-Quantum Connections for the Harmonic Oscillator}

The formalism of quantum mechanics is very different from that of classical
physics. \ However, in some cases, there are striking parallels between the
theories. \ In 1925, Dirac\cite{Dirac} suggested that the transition from
classical to quantum theory could be carried out by replacing the Poisson
brackets of classical mechanics by commutation relations for operators with a
scale given by Planck's constant $\hbar$. \ This classical-quantum
analogy,\cite{PB} taken together with operator symmetrization in the
Hamiltonian, assures that the algebraic structure involving the Hamiltonian
operator in quantum theory matches the algebraic structure associated with the
Hamiltonian function in classical mechanics for many systems.\cite{Dahl} \ For
familiar mechanical systems, such a replacement indeed produces the current
quantum theory, and the Dirac classical-quantum analogy is sometimes mentioned
in the textbooks.\cite{PB} \ However, the connections between the equilibrium
average values of physical quantities in classical and quantum physics involve
aspects different from Dirac's analogy. \ Quantum operators act on Hilbert
space in such a way as to determine the average values of physical quantities
in the ground state. \ In classical mechanics, the equilibrium average values
are determined by integrations over phase space; however, nonrelativistic
classical physics contains no fundamental scale on phase space. \ Thus any
attempt to compare quantum average values with those of some related classical
mechanical system requires an extension beyond nonrelativistic classical mechanics.

The classical extension chosen in this article involves the equilibrium of the
classical mechanical system in classical electromagnetic thermal radiation.
\ Planck's constant $\hbar$ has a natural place within classical
electromagnetic theory as the scale of Lorentz-invariant random classical
electromagnetic radiation, classical zero-point radiation.\cite{B2018a} \ The
constant $\hbar$ is then reflected in the spectrum of equilibrium classical
thermal radiation giving the Planck spectrum with classical electromagnetic
zero-point radiation.\cite{B2018c} \ Finally, the constant $\hbar$ is
reflected in the equilibrium phase space for any mechanical system in
equilibrium with thermal radiation. \ The mechanical system together with the
thermal-equilibrium phase space gives a stochastic classical mechanical
theory. \ In this article, we survey the connections (both similarities and
differences) for harmonic oscillator systems in one and in three spatial
dimensions when described by quantum mechanics or by classical stochastic
mechanics. \ The analysis give us a deeper appreciation of both classical and
quantum theories. \ 

\section{ Algebraic Structure: Commutators vs Poisson Brackets}

\subsection{Fundamental Commutator and Poisson Brackets}

Quantum commutators appear in a student's first course in quantum
mechanics.\cite{G43} \ The easiest form of quantum mechanics seems to involve
the Schroedinger wave function $\psi(x,t)$ for a system of one degree of
freedom. \ In this form, the fundamental quantum commutator for the position
operator $\hat{x}$ and momentum operator $\hat{p}_{x},$
\begin{equation}
\left[  \hat{x},\hat{p}_{x}\right]  =\hat{x}\hat{p}_{x}-\hat{p}_{x}\hat
{x}=i\hbar, \label{F1}%
\end{equation}
can be regarded as the operators $x$ and ($\hbar/i)\partial/\partial x$ acting
on the function space formed by the wave functions $\psi(x,t)$%
\begin{equation}
\left[  \hat{x},\hat{p}_{x}\right]  \psi(x,t)=x\frac{\hbar}{i}\frac{\partial
}{\partial x}\psi(x,t)-\frac{\hbar}{i}\frac{\partial}{\partial x}\left[
x\psi(x,t)\right]  =i\hbar\psi(x,t). \label{F}%
\end{equation}
On the other hand, classical Poisson brackets, which can be used as the basis
for the Hamiltonian formulation of classical mechanics, appear in a graduate
course in classical mechanics.\cite{Goldstein397} \ The classical mechanical
description of a system of one degree of freedom involves canonical variables
$x$ and $p_{x}$ satisfying the fundamental Poisson bracket relation%
\begin{equation}
\left\{  x,p_{x}\right\}  =\frac{\partial x}{\partial x}\frac{\partial p_{x}%
}{\partial p_{x}}-\frac{\partial x}{\partial p_{x}}\frac{\partial p_{x}%
}{\partial x}=1,
\end{equation}
where for any two functions $f(x,p_{x})$ and $g(x,p_{x})$ the Poisson bracket
is%
\begin{equation}
\left\{  f,g\right\}  =\frac{\partial f}{\partial x}\frac{\partial g}{\partial
p_{x}}-\frac{\partial f}{\partial p_{x}}\frac{\partial g}{\partial x}.
\end{equation}
The classical mechanical system is described by a Hamiltonian function
involving the canonical coordinate $x$ and canonical momentum $p_{x},$ and the
Poisson brackets lead to the generation of infinitesimal canonical
transformations. \ 

\subsection{One-Dimensional Harmonic Oscillator}

\subsubsection{Classical Oscillator}

The Hamiltonian describing the one-dimensional harmonic oscillator in
classical physics is
\begin{equation}
H=\frac{p_{x}^{2}}{2m}+\frac{1}{2}m\omega_{0}^{2}x^{2}. \label{Hxp}%
\end{equation}
The Hamiltonian is the generator of canonical time translations, and hence
gives Hamilton's equations of motion in the classical case,%
\begin{equation}
\frac{dx}{dt}=\left\{  x,H\right\}  =\frac{\partial x}{\partial x}%
\frac{\partial H}{\partial p_{x}}-\frac{\partial x}{\partial p_{x}}%
\frac{\partial H}{\partial x}=\frac{p_{x}}{m} \label{xt}%
\end{equation}
and
\begin{equation}
\frac{dp_{x}}{dt}=\left\{  p_{x},H\right\}  =\frac{\partial p_{x}}{\partial
x}\frac{\partial H}{\partial p_{x}}-\frac{\partial p_{x}}{\partial p_{x}}%
\frac{\partial H}{\partial x}=-m\omega_{0}^{2}x. \label{pt}%
\end{equation}
The solutions of the equations of motion are
\begin{align}
x\left(  t\right)   &  =x\left(  0\right)  \cos\omega_{0}t+\left(  \frac
{p_{x}\left(  0\right)  }{m\omega_{0}}\right)  \sin\omega_{0}t,\nonumber\\
p_{x}\left(  t\right)   &  =-m\omega_{0}\,x\left(  0\right)  \sin\omega
_{0}t+p_{x}\left(  0\right)  \cos\omega_{0}t.
\end{align}

\subsubsection{Quantum Oscillator}

The quantum mechanical description of the harmonic oscillator in the
Heisenberg picture follows the same pattern as given above for the classical
oscillator. \ The Hamiltonian now involves operators $\hat{x}$ and $\hat{p},$
\begin{equation}
\hat{H}=\frac{\hat{p}_{x}^{2}}{2m}+\frac{1}{2}m\omega_{0}^{2}\hat{x}^{2}.
\label{Hamq}%
\end{equation}
The equations of motion for the operators $\hat{x}$ and $\hat{p}_{x}$ in the
Heisenberg picture are\cite{S50}
\begin{equation}
i\hbar\frac{d\hat{x}}{dt}=\left[  \hat{x},\hat{H}\right]  =\hat{x}\hat{H}%
-\hat{H}\hat{x}=i\hbar\frac{p_{x}}{m}%
\end{equation}
and
\begin{equation}
i\hbar\frac{d\hat{p}_{x}}{dt}=\left[  \hat{p}_{x},\hat{H}\right]  =\hat{p}%
_{x}\hat{H}-\hat{H}\hat{p}_{x}=-i\hbar m\omega_{0}^{2}\hat{x}.
\end{equation}
The solutions of the equations of motion for the Heisenberg operators are
exactly parallel to the classical case,
\begin{align}
\hat{x}\left(  t\right)   &  =\hat{x}\left(  0\right)  \cos\omega_{0}t+\left(
\frac{\hat{p}_{x}\left(  0\right)  }{m\omega_{0}}\right)  \sin\omega
_{0}t,\nonumber\\
\hat{p}_{x}\left(  t\right)   &  =-m\omega_{0}\,\hat{x}\left(  0\right)
\sin\omega_{0}t+\hat{p}_{x}\left(  0\right)  \cos\omega_{0}t.
\end{align}

\subsection{Three-Dimensional Oscillator and Angular Momentum}

In the case of the three-dimensional oscillator, one can discuss the algebraic
structure involving angular momentum in both classical and quantum theories.
\ The Hamiltonian for the three-dimensional isotropic harmonic oscillator
takes the form of three independent harmonic oscillators along the $x,~y,~$and
$z$ axes,
\begin{equation}
H=\sum\nolimits_{i=1}^{3}\left(  \frac{p_{i}^{2}}{2m}+\frac{1}{2}m\omega
_{0}^{2}x_{i}^{2}\right)  ,
\end{equation}
with the classical Hamiltonian involving canonically conjugate variables
$x_{i}$ and $p_{i},$ and the analogous quantum Hamiltonian $\hat{H}$ involving
quantum operators $\hat{x}_{i}$ and $\hat{p}_{i}$.

The $z$-component of angular momentum of the particle in a three-dimensional
harmonic potential involves the function $L_{z}=xp_{y}-yp_{x}$ in classical
physics, and $\hat{L}_{z}=\hat{x}\hat{p}_{y}-\hat{y}\hat{p}_{x}$ in the
quantum formulation. \ The components of angular momentum around the $x$- and
$y$-axis are formed analogously. \ Using Poisson brackets, the angular
momentum functions are generators of the 3-dimensional rotation group. \ The
Poisson brackets now involve $x,~p_{x},~y,~p_{y},~z,$~$p_{z}.$ \ Thus we have%
\begin{align}
\left\{  L_{x},L_{y}\right\}   &  =\frac{\partial L_{x}}{\partial x}%
\frac{\partial L_{y}}{\partial p_{x}}-\frac{\partial L_{x}}{\partial p_{x}%
}\frac{\partial L_{y}}{\partial x}+\frac{\partial L_{x}}{\partial y}%
\frac{\partial L_{y}}{\partial p_{y}}-\frac{\partial L_{x}}{\partial p_{y}%
}\frac{\partial L_{y}}{\partial y}+\frac{\partial L_{x}}{\partial z}%
\frac{\partial L_{y}}{\partial p_{z}}-\frac{\partial L_{x}}{\partial p_{z}%
}\frac{\partial L_{y}}{\partial z}\nonumber\\
&  =\frac{\partial\left(  yp_{z}-zp_{y}\right)  }{\partial z}\frac
{\partial\left(  zp_{x}-xp_{z}\right)  }{\partial p_{z}}-\frac{\partial\left(
yp_{z}-zp_{y}\right)  }{\partial p_{z}}\frac{\partial\left(  zp_{x}%
-xp_{z}\right)  }{\partial z}=xp_{y}-yp_{x}=L_{z}.
\end{align}
In general, we have in classical mechanics%
\begin{equation}
\left\{  L_{i},L_{j}\right\}  =\epsilon_{ijk}L_{k}.
\end{equation}
On the other hand, within quantum physics, the commutator gives%
\begin{align}
\left[  \hat{L}_{x},\hat{L}_{y}\right]   &  =\hat{L}_{x}\hat{L}_{y}-\hat
{L}_{y}L_{x}=\left(  \hat{y}\hat{p}_{z}-\hat{z}\hat{p}_{y}\right)  \left(
\hat{z}\hat{p}_{x}-\hat{x}\hat{p}_{z}\right)  -\left(  \hat{z}\hat{p}_{x}%
-\hat{x}\hat{p}_{z}\right)  \left(  \hat{y}\hat{p}_{z}-z\hat{p}_{y}\right)
\nonumber\\
&  =\hat{y}\hat{p}_{z}\hat{z}\hat{p}_{x}+\hat{z}\hat{p}_{y}\hat{x}\hat{p}%
_{z}-\hat{z}\hat{p}_{x}\hat{y}\hat{p}_{z}-\hat{x}\hat{p}_{z}\hat{z}\hat{p}%
_{Y}\nonumber\\
&  =\hat{y}\left[  \hat{p}_{z},\hat{z}\right]  \hat{p}_{x}-\hat{x}\left[
\hat{p}_{z},\hat{z}\right]  \hat{p}_{y}=i\hbar(\hat{x}\hat{p}_{y}-\hat{y}%
\hat{p}_{x})=i\hbar\hat{L}_{z}.
\end{align}
In general, we have in quantum mechanics%
\begin{equation}
\left[  \hat{L}_{i},\hat{L}_{j}\right]  =\epsilon_{ijk}i\hbar\hat{L}_{k}.
\end{equation}
Thus in both the classical and quantum theories the angular momentum is
associated with the algebra of the infinitesimal generators of the rotation
group. \ 

In addition, it is easy to prove that the square of the angular momentum
$L^{2}=L_{x}^{2}+L_{y}^{2}+L_{z}^{2}$ satisfies
\begin{equation}
\left\{  L^{2},L_{i}\right\}  =0
\end{equation}
for the classical Poisson brackets, and that the quantum operator $\hat{L}%
^{2}=\hat{L}_{x}^{2}+\hat{L}_{y}^{2}+\hat{L}_{z}^{2}$ commutes with all the
angular momentum operators%
\begin{equation}
\left[  \hat{L}^{2},\hat{L}_{i}\right]  =0.
\end{equation}
Thus the square of the angular momentum acts as a Casimir operator in the Lie
algebra of the generators of the three-dimensional rotation
group.\cite{Zee210} \ The values taken by the Casimir operator $L^{2}$
correspond to $l\left(  l+1\right)  $ in the classical theory, and $\hbar
^{2}l\left(  l+1\right)  $ in the quantum theory. \ 

\section{Numerical Values: Hilbert Space vs Phase Space}

\subsection{Average Values in Classical and Quantum Mechanics}

Dirac emphasized that the association of the Poisson bracket with the
commutation would give parallel algebraic structures in classical and quantum
theories.\cite{Dirac} \ However, the theories involve more than simply their
algebraic structure. \ In order to obtain equilibrium numerical values for
physical quantities, the classical theory must evaluate the canonical
variables over the appropriate phase space probability distribution. \ For
motion in one spatial dimension, the phase space distribution $P(x,p_{x})$
gives the average value of the quantity $f(x,p_{x})$ as%
\begin{equation}
\left\langle f\left(  x,p_{x}\right)  \right\rangle _{classical}%
=\int\nolimits_{-\infty}^{\infty}dx\int\nolimits_{-\infty}^{\infty}%
dp_{x}\,f(x,p_{x})P\left(  x,p_{x}\right)  .
\end{equation}
Such an evaluation is familiar from classical statistical mechanics where the
phase space distribution corresponds to that of thermal equilibrium. \ If the
classical phase space involves simply fixed initial conditions, the
probability distribution reduces to a single point on the phase space.

On the other hand, the quantum theory at zero temperature obtains average
values by applying the quantum operators to Hilbert space as the expectation
value
\begin{equation}
\left\langle f(\hat{x},\hat{p}_{x})\right\rangle _{quantum}=\int%
\nolimits_{-\infty}^{\infty}dx\,\psi^{\ast}\left(  x,t\right)  f(x,\frac
{\hbar}{i}\frac{\partial}{\partial x})\psi(x,t).
\end{equation}
We notice that quantum theory contains the constant $\hbar$ attached to the
fundamental commutator in \ Eq. (\ref{F}), and this constant will serve as a
scale for all the operator expectation values evaluated on the Hilbert space.
\ For example, if we introduce the raising and lowering operators $\hat{a}$
and $\hat{a}^{+}$ where\cite{G44}
\begin{equation}
\hat{a}=\sqrt{\frac{m\omega_{0}}{2\hbar}}\left(  \hat{x}+\frac{i\hat{p}_{x}%
}{m\omega_{0}}\right)  \text{ \ and \ }\hat{a}^{+}=\sqrt{\frac{m\omega_{0}%
}{2\hbar}}\left(  \hat{x}-\frac{i\hat{p}_{x}}{m\omega_{0}}\right)  ,
\end{equation}
then the quantum Hamiltonian in Eq. (\ref{Hamq}) can be rewritten as
\begin{equation}
H=\hbar\omega_{0}\left(  \hat{a}^{+}\hat{a}+\frac{1}{2}\right)  . \label{Ha}%
\end{equation}
\ One can show\cite{G46} that the operator $\hat{a}^{+}$ acting on the vacuum
state can generate states in Hilbert space with energies ($n+1/2)\hbar
\omega_{0}$ where $n=0,1,2,...$

The average energy of the classical oscillator is obtained by evaluating the
values of position and momentum over the phase space which describes the
classical system. \ However, nonrelativistic classical mechanics has nothing
to say about a \textit{fundamental} phase space distribution $P(x,p_{x}).$

\subsection{Traditional Contrast Between Classical and Quantum Theories\ }

\ It is exactly at this point that the traditional treatments of classical
theory and quantum theory part company. \ Planck's constant $\hbar$ does not
appear in the classical mechanical algebra of infinitesimal canonical
transformations, and there is no fundamental role for the constant in
traditional nonrelativistic classical mechanical phase space distributions.
\ Thus the traditional claim arises that Planck's constant $\hbar$ appears in
quantum theory, but not in classical theory. \ Indeed, Goldstein's classical
mechanics text book points out the SO(4) algebraic symmetry of a classical
particle in a Coulomb or Kepler potential,\cite{Goldstein421} but the
classical mechanics text book goes no further, while Pauli\cite{Pauli}
exploited this same symmetry in quantum theory in order to derive the Balmer
spectrum for hydrogen in 1926.

There is a striking contrast between the classical and quantum systems.
\ \textit{The quantum system contains a scale parameter }$\hbar$%
\textit{\ attached to the operators of the theory which arises from the
fundamental operator commutator. \ This scale will produce the ground state
energy value for the quantum harmonic oscillator on Hilbert space.} \ The
classical mechanical situation is completely different. \ \textit{The
classical mechanical system contains no scale within the canonical
transformation formalism produced by the Poisson brackets. \ The values of
quantities in the classical mechanical system follow from the assumed phase
space distribution.}

\section{Phase Space in Classical Theory with Classical Electromagnetic
Zero-Point Radiation}

\subsection{A Closer Classical Approximation to Quantum Physics}

We are interested in the connections between classical and quantum theories
when the classical theory is not simply nonrelativistic classical mechanics
with no scale $\hbar,$ but rather is the classical theory which comes as close
as possible to approximating quantum theory. \ We must go outside
nonrelativistic classical mechanic in order to accomplish this aim. \ One
possible extension turns to classical electromagnetic theory.\cite{SED}

The arguments leading to an extension into classical electromagnetic theory
are as follows. \ Small bits of matter in a fluid at room temperature are
observed to perform a random Brownian motion, presumably forced into motion by
collisions with the invisible molecules of the solution. \ It is therefore
assumed that, in equilibrium, a small harmonic oscillator would be forced into
thermal motion by colliding gas molecules. \ Even if all the gas molecules
were removed, a charged harmonic oscillator would still come to thermal
equilibrium with the ambient thermal radiation. \ Indeed, Planck's
calculation\cite{Planck} at the end of the 19th century showed that a
classical charged harmonic oscillator acquires an average energy equal to the
average energy per normal mode of the surround random classical radiation at
the frequency $\omega_{0}$ of the oscillator. \ In the limit as the
temperature is decreased to zero, the charged harmonic oscillator would be in
equilibrium with the random radiation which exists at zero temperature. \ This
random radiation at the zero of temperature is termed classical
electromagnetic zero-point radiation. \ The Casimir force\cite{Casimir}
between two uncharged parallel conducting plates depends upon all the
radiation surrounding the plates, and experimental measurements\cite{exp} show
that at zero temperature, the Casimir force does not vanish. \ Rather, the
Casimir force can be explained quantitatively by the existence of random
classical radiation with an energy spectrum $\mathcal{E}_{\omega}=\left(
1/2\right)  \hbar\omega$ per normal mode.\cite{B1973} \ Therefore within a
purely classical theory, one must assume that all matter exists in ambient
classical electromagnetic thermal radiation which enforces an equilibrium
phase space distribution on any mechanical system. \ At zero temperature,
classical thermal radiation is classical zero-point radiation with a
Lorentz-invariant spectrum and a scale set by Planck's constant $\hbar.$ \ The
existence of classical zero-point radiation at zero temperature is sufficient
to allow a classical derivation of the full Planck radiation spectrum
(including zero-point radiation) at positive temperature.\cite{B2018c} \ 

The behavior of a classical mechanical system, such as a harmonic oscillator,
which interacts with classical thermal radiation will come to equilibrium with
a phase space distribution which reflects the randomness of the thermal
radiation. \ Even if one takes the limit as the interaction between the
oscillator and the thermal radiation becomes ever smaller, the phase space
distribution for the mechanical system will remain unchanged, though the time
required to reach equilibrium will increase as the interaction strength
decreases. \ Indeed this situation is familiar for systems in thermal
equilibrium where the coupling between thermodynamic systems can be made
arbitrarily small and yet the equilibrium phase space distribution for fixed
temperature $T$ remains the same, independent of the (small) size of the
inter-system coupling. \ Thus in equilibrium with classical thermal radiation,
any classical mechanical system will acquire a phase space distribution which
reflects the random character of the thermal radiation and contains the scale
factor $\hbar$ arising from the scale of classical electromagnetic zero-point radiation.

\subsection{Classical Phase Space from Classical Thermal Radiation Including
Zero-Point Radiation}

In equilibrium, the equation of motion for the harmonic oscillator must
include driving by classical thermal radiation with the Planck spectrum
including zero-point radiation, corresponding to an energy per normal mode
$\mathcal{E}(\omega,T)=\left(  \hbar\omega_{0}/2\right)  \coth\left[
\hbar\omega_{0}/\left(  k_{B}T\right)  \right]  =\hbar\omega\lbrack\exp
(\hbar\omega/k_{B}T)-1]^{-1}+\hbar\omega/2$. \ Thus the correct classical
equation of motion obtained by combining equations (\ref{xt}) and (\ref{pt})
is not the purely mechanical expression $m\ddot{x}=-m\omega_{0}^{2}x$ but
rather must be written as a Langevin equation involving fluctuation and
damping. \ The oscillating particle is assumed to carry a (small) charge $e$
so that it interacts with random thermal radiation as\cite{B1975}
\begin{equation}
m\ddot{x}=-m\omega_{0}^{2}x+m\tau\dddot{x}+eE_{x}^{T}(x,t), \label{xeqt}%
\end{equation}
where $\tau=(2e^{2})/(3mc^{3})$ is associated with radiation damping, and
$eE_{x}^{T}(x,t)$ is the force on the oscillator due to the random classical
thermal radiation. \ The equation of motion (\ref{xeqt}) has been solved (for
the steady-state solution) many times\cite{many} going back to Planck's
work.\cite{Planck} \ The result for the equilibrium phase space distribution
$P_{T}(x,p_{x})$ for the oscillator in random classical radiation
(corresponding to the Planck spectrum at temperature $T$ and including
zero-point radiation) is independent of the (small) charge $e$ and takes the
form%
\begin{equation}
P_{T}(x,p_{x})dxdp_{x}=\frac{1}{2\pi\hbar\coth\left[  \hbar\omega_{0}/\left(
k_{B}T\right)  \right]  }\exp\left(  -\frac{p_{x}^{2}/2m+m\omega_{0}^{2}%
x^{2}/2}{\left(  \hbar\omega_{0}/2\right)  \coth\left[  \hbar\omega
_{0}/\left(  k_{B}T\right)  \right]  }\right)  dxdp_{x} \label{PTxp}%
\end{equation}
for $-\infty<x<\infty,~-\infty<p_{x}<\infty.$ \ At zero temperature
$T\rightarrow0$, the phase space distribution becomes%
\begin{equation}
P_{0}(x,p_{x})dxdp_{x}=\frac{1}{2\pi\hbar}\exp\left(  -\frac{p_{x}%
^{2}/2m+m\omega_{0}^{2}x^{2}/2}{\left(  \hbar\omega_{0}/2\right)  }\right)
dxdp_{x}. \label{P0xp}%
\end{equation}
\ Thus at zero temperature, the scale $\hbar$ of the classical zero-point
radiation which appears in the energy $\hbar\omega/2~$per radiation normal
mode reappears as the scale factor for the oscillator on phase space. \ The
phase space distribution in Eq. (\ref{P0xp}) is a probability distribution on
phase space for the position and momentum of the particle under the influence
of classical zero-point radiation. \ Thus in classical physics with classical
electromagnetic zero-point radiation, the canonical transformations determined
by the Poisson brackets are separate from the scale of the phase space
distribution determined by the oscillator's interaction with classical thermal
radiation. \ 

\section{Comparison of Ground States for the Classical and Quantum
Oscillators}

\subsection{One-Dimensional Oscillator}

Since we now have an example of a classical phase space, we can compare and
contrast the ground state descriptions for the harmonic oscillator given by
the classical and quantum theories. \ The quantum description is by far the
more familiar. \ The quantum ground state wave function is given by\cite{G46}
\begin{equation}
\psi_{0}(x)=\left(  \frac{m\omega_{0}}{\pi\hbar}\right)  ^{1/4}\exp\left(
-\frac{m\omega_{0}x^{2}}{2\hbar}\right)  . \label{psi0}%
\end{equation}
It is worth noting that the square of the quantum wave function in Eq.
(\ref{psi0}) agrees exactly with the classical phase space distribution for
$x$ independent of $p_{x}$ found by integrating over the classical phase space
distribution in Eq. (\ref{P0xp}) with respect to $p_{x}$ so as to remove the
reference to the momentum $p_{x}$,%
\begin{equation}
|\psi_{0}(x)|^{2}=\left(  \frac{m\omega_{0}}{\pi\hbar}\right)  ^{1/2}%
\exp\left(  -\frac{m\omega_{0}x^{2}}{\hbar}\right)  =\int\nolimits_{-\infty
}^{\infty}dp_{x}P_{0}(x,p_{x}).
\end{equation}
\ The average value for the $n$th power of the position of the particle is
given by
\begin{equation}
\left\langle \hat{x}^{n}\right\rangle _{0\text{~}quantum}=\int%
\nolimits_{-\infty}^{\infty}dx\psi_{0}^{\ast}x^{n}\psi_{0}=\frac{\left(
2n\right)  !}{\left(  n!\right)  ^{2}2^{n}}\left(  \frac{\hbar}{2m\omega_{0}%
}\right)  ,
\end{equation}
which is the same as the average value for the classical oscillator over the
classical phase space distribution in Eq. (\ref{P0xp}) arising from the
zero-point radiation%
\begin{equation}
\left\langle x^{n}\right\rangle _{0~classical}=\int\nolimits_{-\infty}%
^{\infty}dp_{x}\int\nolimits_{-\infty}^{\infty}dx\,x^{n}P_{0}(x,p_{x}%
)=\frac{\left(  2n\right)  !}{\left(  n!\right)  ^{2}2^{n}}\left(  \frac
{\hbar}{2m\omega_{0}}\right)  . \label{xncl}%
\end{equation}
Similarly, the average values for the powers of momenta agree. \ Thus the
average value for $\hat{p}_{x}^{n}$ is
\begin{equation}
\left\langle \hat{p}_{x}^{n}\right\rangle _{0~quantum}=\int\nolimits_{-\infty
}^{\infty}dx\psi_{0}^{\ast}\left(  \frac{\hbar}{i}\frac{d}{dx}\right)
^{n}\psi_{0}=\frac{\left(  2n\right)  !}{\left(  n!\right)  ^{2}2^{n}}\left(
\frac{\hbar m\omega_{0}}{2}\right)  ,
\end{equation}
which is the same as the average value for the classical oscillator over the
phase space distribution arising from the zero-point radiation,%
\begin{equation}
\left\langle p_{x}^{n}\right\rangle _{0~classical}=\int\nolimits_{-\infty
}^{\infty}dp_{x}\int\nolimits_{-\infty}^{\infty}dx\,p_{x}^{n}P_{0}%
(x,p_{x})=\frac{\left(  2n\right)  !}{\left(  n!\right)  ^{2}2^{n}}\left(
\frac{\hbar m\omega_{0}}{2}\right)  . \label{pncl}%
\end{equation}

Because both theories agree on the average values $\left\langle \hat{x}%
^{2}\right\rangle =$ $\left\langle x^{2}\right\rangle $ and $\left\langle
\hat{p}_{x}^{2}\right\rangle =\left\langle p_{x}^{2}\right\rangle ,$ both
theories agree on the average value of energy,%
\begin{equation}
\left\langle \hat{H}\right\rangle _{0~quantum}=\int\nolimits_{-\infty}%
^{\infty}dx\psi_{0}^{\ast}H\psi_{0}=\frac{\left\langle \hat{p}_{x}%
^{2}\right\rangle }{2m}+\frac{m\omega_{0}^{2}\left\langle \hat{x}%
^{2}\right\rangle }{2}=\frac{1}{2}\hbar\omega_{0}, \label{Eqq}%
\end{equation}
and
\begin{equation}
\left\langle H\right\rangle _{0~classical}=\int\nolimits_{-\infty}^{\infty
}dp_{x}\int\nolimits_{-\infty}^{\infty}dx\,Hf(x,p_{x})=\frac{\left\langle
p_{x}^{2}\right\rangle }{2m}+\frac{m\omega_{0}^{2}\left\langle x^{2}%
\right\rangle }{2}=\frac{1}{2}\hbar\omega_{0}.
\end{equation}

However, the theories differ regarding the fluctuations of energy. \ For
example, the quantum theory involves an energy eigenstate and involves no
fluctuations in energy, so that we find%
\begin{align}
\left\langle \hat{H}^{2}\right\rangle _{0~quantum}  &  =\int\nolimits_{-\infty
}^{\infty}dx\psi_{0}^{\ast}H^{2}\psi_{0}=\left\langle \left(  \frac{\hat
{p}_{x}^{2}}{2m}+\frac{1}{2}m\omega_{0}^{2}\hat{x}^{2}\right)  ^{2}%
\right\rangle \nonumber\\
&  =\frac{\left\langle \hat{p}_{x}^{4}\right\rangle }{\left(  2m\right)  ^{2}%
}+\frac{\omega_{0}^{2}}{4}\left(  \left\langle \hat{p}_{x}^{2}\hat{x}%
^{2}\right\rangle +\left\langle \hat{x}^{2}\hat{p}_{x}^{2}\right\rangle
\right)  +\left(  \frac{m\omega_{0}^{2}}{2}\right)  ^{2}\left\langle \hat
{x}^{4}\right\rangle \nonumber\\
&  =\left(  \frac{1}{2}\hbar\omega_{0}\right)  ^{2}=\left\langle \hat
{H}\right\rangle _{0~quantum}^{2}, \label{H2q}%
\end{align}
since the quantum operators give $\left\langle \hat{x}^{2}\hat{p}_{x}%
^{2}\right\rangle =\left\langle \hat{p}_{x}^{2}\hat{x}^{2}\right\rangle
=-\left\langle \hat{x}^{2}\right\rangle \left\langle \hat{p}_{x}%
^{2}\right\rangle .$ \ On the other hand, the energy of the classical
oscillator indeed fluctuates giving
\begin{align}
\left\langle H^{2}\right\rangle _{0~classical}  &  =\int\nolimits_{-\infty
}^{\infty}dp_{x}\int\nolimits_{-\infty}^{\infty}dx\,H^{2}P_{0}(x,p_{x}%
)\nonumber\\
&  =\frac{\left\langle p_{x}^{4}\right\rangle }{\left(  2m\right)  ^{2}}%
+\frac{\omega_{0}^{2}}{2}\left\langle p_{x}^{2}x^{2}\right\rangle +\left(
\frac{m\omega_{0}^{2}}{2}\right)  ^{2}\left\langle x^{4}\right\rangle
=2\left\langle H\right\rangle _{0~classical}^{2},
\end{align}
since for the classical quantities $\left\langle x^{2}p_{x}^{2}\right\rangle
=\left\langle x^{2}\right\rangle \left\langle p_{x}^{2}\right\rangle .$ \ The
classical and quantum expressions for $\left\langle \hat{H}^{2}\right\rangle
_{0~quantum}$ and $\left\langle H^{2}\right\rangle _{0~classical}$ do not
agree. \ In the quantum mechanical ground state, the position and momentum of
the particle fluctuate, but the fluctuations are correlated in a special
non-classical way\cite{gencon} such that the energy takes a discrete value;
the energy of the quantum oscillator is an eigenvalue for the ground state
energy. \ In contrast, the probabilities of the classical system are fully
described by the probability distribution on phase space. \ For the classical
oscillator at zero temperature, the randomness is derived from the randomness
of classical zero-point radiation. The energy of the system in equilibrium is
a slowly varying quantity which explores the phase space as the oscillator
exchanges energy with the classical zero-point radiation, and the energy
probability distribution can be obtained from the phase space
distribution.\cite{HBsim} \ \ The disagreement between the fluctuations of the
theories has been known for some time. \ Whether the distinction is
experimentally measurable is another matter. \ So far as I know, no one has
viewed the disagreement as appropriate for a possible experimental measurement.

\subsection{Three-Dimensional Harmonic Oscillator}

The agreement and the contrasts between the quantum theory and the classical
theory with zero-point radiation can be illustrated further by considering the
three-dimensional harmonic oscillator. \ For this situation, we find
interesting contrasts in the treatment of angular momentum. \ 

The Hamiltonian for the three-dimensional isotropic harmonic oscillator takes
the form of three independent harmonic oscillators along the $x,~y,~$and $z$
axes,
\begin{equation}
H=\sum\nolimits_{i=1}^{3}\left(  \frac{p_{i}^{2}}{2m}+\frac{1}{2}m\omega
_{0}^{2}x_{i}^{2}\right)  ,
\end{equation}
with the classical Hamiltonian involving canonically conjugate variables
$x_{i}$ and $p_{i},$ and the quantum Hamiltonian $\hat{H}$ involving quantum
operators $\hat{x}_{i}$ and $\hat{p}_{i}$. \ In the ground state at zero
temperature, the classical distribution on phase space involves a product
probability distribution%
\begin{equation}
P_{0}(x,p_{x},y,p_{y}z,p_{z})=\prod\nolimits_{i=1}^{3}\frac{1}{\left(
\pi\hbar\right)  ^{3}}\exp\left(  -\frac{p_{i}^{2}/(2m)+(1/2)m\omega_{0}%
^{2}x_{i}^{2}}{\hbar\omega_{0}/2}\right)  , \label{ph3}%
\end{equation}
and the quantum wave function is given as a product state in Hilbert space%
\begin{equation}
\psi(x,y,z)=\prod\nolimits_{i=1}^{3}\left(  \frac{m\omega_{0}}{\pi\hbar
}\right)  ^{3/4}\exp\left(  -\frac{m\omega_{0}x_{i}^{2}}{2\hbar}\right)  .
\label{psi3}%
\end{equation}
Once again, in the ground state, the quantum wave function squared corresponds
to the classical phase space distribution integrated over the momentum variables%

\begin{equation}
|\psi(x,y,z)|^{2}=\int dp_{x}dp_{y}dp_{z}P_{0}(x,p_{x},y,p_{y}z,p_{z}).
\end{equation}

Now the classical ground-state phase space distribution in Eq. (\ref{ph3}) and
the quantum ground-state wave function in Eq. (\ref{psi3}) are both
rotationally invariant and so correspond to the identity representation of the
rotation group. \ Indeed, applying the Poisson bracket to find $L^{2}$ acting
on the classical ground state phase space distribution $P_{0}(x,p_{x}%
,y,p_{y},z,p_{z})$, we have
\begin{equation}
L^{2}=(yp_{z}-zp_{y})^{2}+\left(  zp_{x}-xp_{z}\right)  ^{2}+\left(
xp_{y}-yp_{x}\right)  ^{2},
\end{equation}%
\begin{align}
\frac{\partial\left(  L^{2}\right)  }{\partial x}  &  =-2p_{z}(zp_{x}%
-xp_{z})+2p_{y}\left(  xp_{y}-yp_{x}\right)  =\nonumber\\
&  =2x\left(  p_{y}^{2}+p_{z}^{2}\right)  -2p_{x}(zp_{z}+yp_{y}),
\end{align}%
\begin{align}
\frac{\partial\left(  L^{2}\right)  }{\partial p_{x}}  &  =2z\left(
zp_{x}-xp_{z}\right)  -2y\left(  xp_{y}-yp_{x}\right) \nonumber\\
&  =+2p_{x}(y^{2}+z^{2})-2x(yp_{y}+zp_{z}),
\end{align}
while from Eq. (\ref{ph3}),
\begin{equation}
\frac{\partial P_{0}}{\partial x}=m\omega_{0}^{2}xP_{0}\text{\ and }%
\frac{\partial P_{0}}{\partial p_{x}}=\frac{p_{x}}{m}P_{0,}%
\end{equation}
so that%
\begin{equation}
\left\{  P_{0},L^{2}\right\}  =-\left(  \frac{\partial L^{2}}{\partial x}%
\frac{\partial P_{0}}{\partial p_{x}}-\frac{\partial L^{2}}{\partial p_{x}%
}\frac{\partial P_{0}}{\partial x}+\frac{\partial L^{2}}{\partial y}%
\frac{\partial P_{0}}{\partial p_{y}}-\frac{\partial L^{2}}{\partial p_{y}%
}\frac{\partial P_{0}}{\partial y}+\frac{\partial L^{2}}{\partial z}%
\frac{\partial P_{0}}{\partial p_{z}}-\frac{\partial L^{2}}{\partial p_{z}%
}\frac{\partial P_{0}}{\partial z}\right)  =0. \label{L2P0}%
\end{equation}
Also, if we have the quantum operator $\hat{L}^{2}$ act on the ground state
wave function $\psi(x,y,z)$ in Eq. (\ref{psi3}) using
\begin{equation}
\hat{L}^{2}\psi_{0}(x,y,z)=-\hbar^{2}\left\{  \left(  y\frac{\partial
}{\partial z}-z\frac{\partial}{\partial y}\right)  ^{2}+\left(  z\frac
{\partial}{\partial x}-x\frac{\partial}{\partial z}\right)  ^{2}+\left(
x\frac{\partial}{\partial y}-y\frac{\partial}{\partial x}\right)
^{2}\right\}  \psi(x,y,z),
\end{equation}
we find%
\begin{equation}
\hat{L}^{2}\psi_{0}(x,y,z)=0. \label{psiL2}%
\end{equation}
Thus indeed, the ground states provides a basis for the identity
representation of the three-dimensional rotation algebra corresponding to
$l=0.$

\subsection{Interpretation of the Physical Angular Momentum Squared in the
Ground State}

The spherically-symmetric ground state of the three-dimensional harmonic
oscillator is invariant under all the infinitesimal generators of rotation.
\ It also corresponds to a Casimir operator value of $l=0.$ \ However, just as
the quantum theory and classical theory differ in their interpretation of the
fluctuations in the energy of the ground state, the classical and quantum
theories differ in their interpretations of the physical angular momentum.
\ In the quantum theory, the hermitian operator $\hat{L}^{2}$ is connected to
the physical angular momentum, and the average value for the square of the
angular momentum of the system in the ground state follows from Eq.
(\ref{psiL2}) as
\begin{equation}
\left\langle \hat{L}^{2}\right\rangle _{0~quantum}=\int\nolimits_{-\infty
}^{\infty}d^{3}x\psi_{0}^{\ast}\hat{L}^{2}\psi_{0}=0. \label{avqL2}%
\end{equation}
On the other hand, in the classical theory, the numerical value for the
angular momentum squared is not obtained from the algebra of the Poisson
bracket acting on phase space as in Eq. (\ref{L2P0}), but rather is obtained
by averaging over the ground state phase space distribution for the harmonic
oscillator. \ We notice that
\begin{align}
\left\langle L_{x}^{2}\right\rangle _{0~classical}  &  =\int\nolimits_{-\infty
}^{\infty}d^{3}p\int\nolimits_{-\infty}^{\infty}d^{3}x\,(yp_{z}-zp_{y}%
)^{2}P_{0}(x,p_{x},y,p_{y}z,p_{z})\nonumber\\
&  =\int\nolimits_{-\infty}^{\infty}d^{3}p\int\nolimits_{-\infty}^{\infty
}d^{3}x\,(y^{2}p_{z}^{2}+z^{2}p_{y}^{2}-2yzp_{y}p_{z})P_{0}(x,p_{x}%
,y,p_{y}z,p_{z})\nonumber\\
&  =2\left(  \frac{\hbar}{2m\omega_{0}}\right)  \left(  \frac{\hbar
m\omega_{0}}{2}\right)  =\frac{\hbar^{2}}{2},
\end{align}
with the same values for $\left\langle L_{y}^{2}\right\rangle _{0~classical}%
$\ and $\left\langle L_{z}^{2}\right\rangle _{0~classical}$\ .
\ \ Accordingly, we find%
\begin{equation}
\left\langle L^{2}\right\rangle _{0~classical}=\left\langle L_{x}^{2}%
+L_{y}^{2}+L_{z}^{2}\right\rangle _{0~classical}=\int\nolimits_{-\infty
}^{\infty}d^{3}p\int\nolimits_{-\infty}^{\infty}d^{3}x\,L^{2}P_{0}%
(x,p_{x},y,p_{y}z,p_{z})=\frac{3}{2}\hbar^{2}. \label{avclL2}%
\end{equation}
Thus the algebraic aspects associated with the representations of the rotation
group appearing in Eqs. (\ref{L2P0}) and (\ref{psiL2}) agree between the
classical and quantum theories. However, the average values for the square of
the angular momentum of the physical system are given in Eqs. (\ref{avqL2})
and (\ref{avclL2}), and these do not agree between the classical and quantum
theories. \ In the physical interpretation based upon the classical theory
with zero-point radiation, the angular momentum of the particle in the ground
states changes gradually due to interaction with the random zero-point
radiation; sometimes the particle is moving around the potential center
thereby giving a non-zero value for $\left\langle L^{2}\right\rangle
_{0~classical}$. \ Thus in the classical interpretation where the phase space
distribution arises from the random classical zero-point radiation, the mean
square of the angular momentum does not vanish. \ On the other hand, the
quantum result $\left\langle \hat{L}^{2}\right\rangle _{0~quantum}=0$
(involving the vanishing square of the angular momentum in the quantum ground
state) is sometimes given a semiclassical interpretation with the suggestion
that the particle is always moving directly toward the potential center or
else directly away, but never going around the potential center. \ Whether
this distinction is experimentally measurable at the microscopic level is not
obvious. \ 

\section{Classical and Quantum Theories at Positive Temperature}

\subsection{Continuity of the Classical Analysis}

It is interesting to see the classical-quantum connections extended to
equilibrium situations involving positive temperature. \ The classical theory
makes no change in the conceptual framework as the temperature is increased
above absolute zero. \ Thus at positive temperature, the classical phase space
distribution merely changes from Eq. (\ref{P0xp}) over to Eq. (\ref{PTxp})
because the thermal radiation has changed. \ However, the classical mechanical
phase space distribution at temperature $T>0$ still corresponds to the
equilibrium situation in the thermal radiation corresponding to Planck's
spectrum including zero-point radiation. \ For the one-dimensional harmonic
oscillator, Eqs. (\ref{Hxp}) and (\ref{PTxp}) give
\begin{equation}
\left\langle H\right\rangle _{T~classical}=\int dxdp_{x}\,H\,P_{T}%
(x,p_{x})=\frac{1}{2}\hbar\omega_{0}\coth\left(  \frac{\hbar\omega_{0}}%
{2k_{B}T}\right)  .
\end{equation}
The average value of the classical mechanical oscillator energy changes
continuously from the value at zero temperature because the phase space
distribution has changed continuously. \ The fluctuations in energy again
follow from the classical phase space probability distribution. \ Thus we
find
\begin{equation}
\left\langle H^{2}\right\rangle _{T~classical}=\int dxdp_{x}\,H^{2}%
\,P_{T}(x,p_{x})=2\left(  \frac{1}{2}\hbar\omega_{0}\coth\left(  \frac
{\hbar\omega_{0}}{2k_{B}T}\right)  \right)  ^{2}=2\left\langle H\right\rangle
_{T~classical}^{2}.
\end{equation}

\subsection{Change in the Quantum Analysis}

In contrast to the classical situation, the quantum situation undergoes a
significant change. \ The quantum state can no longer be characterized by a
single vector on the Hilbert space. \ Rather, the expectation value for a
quantum operator at positive temperature is given as a sum over the
expectation values in all the excited states with a weighting corresponding to
the Boltzmann factor. \ For example, the expectation value of the energy is
given by
\begin{equation}
\left\langle \hat{H}\right\rangle _{T~quantum}=\sum\nolimits_{i}\left(
\int\nolimits_{-\infty}^{\infty}d^{3}x\psi_{i}^{\ast}\hat{H}\psi_{i}\right)
\frac{\exp\left[  -\mathcal{E}_{i}/\left(  k_{B}T\right)  \right]  }{Z},
\end{equation}
where the probability normalization requires the normalization factor%
\begin{equation}
Z=\sum\nolimits_{i}\exp\left[  -\mathcal{E}_{i}/\left(  k_{B}T\right)
\right]  . \label{Z}%
\end{equation}
For the case of the one-dimensional harmonic oscillator, the sum in Eq.
(\ref{Z}) becomes a geometric series giving
\begin{equation}
Z=\frac{1}{2\sinh\left[  \hbar\omega_{0}/\left(  2k_{B}T\right)  \right]  },
\end{equation}
and we find, by taking the derivative of the geometric series with respect to
$\left(  k_{B}T\right)  ^{-1},$%
\begin{align}
\left\langle \hat{H}\right\rangle _{T~quantum}  &  =\frac{1}{Z}\sum
\nolimits_{n=0}^{\infty}\left(  n+\frac{1}{2}\right)  \hbar\omega_{0}%
\exp\left[  -\mathcal{(}n+1/2)\hbar\omega_{0}/\left(  k_{B}T\right)  \right]
\nonumber\\
&  =\frac{1}{2}\hbar\omega_{0}\coth\left(  \frac{\hbar\omega_{0}}{2k_{B}%
T}\right)  .
\end{align}
Thus again at positive temperature, the average value of the energy of the
harmonic oscillator agrees between classical and quantum theories. \ On the
other hand, the quantum energy fluctuations for the oscillator involve%
\begin{align}
\left\langle \hat{H}^{2}\right\rangle _{T~quantum}  &  =\sum\nolimits_{i}%
\left(  \int\nolimits_{-\infty}^{\infty}d^{3}x\psi_{i}^{\ast}\hat{H}^{2}%
\psi_{i}\right)  \frac{\exp\left[  -\mathcal{E}_{i}/\left(  k_{B}T\right)
\right]  }{Z}\nonumber\\
&  =\frac{1}{Z}\sum\nolimits_{n=0}^{\infty}\left[  \left(  n+\frac{1}%
{2}\right)  \hbar\omega_{0}\right]  ^{2}\exp\left[  -\mathcal{(}%
n+1/2)\hbar\omega_{0}/\left(  k_{B}T\right)  \right] \nonumber\\
&  =2\left(  \frac{1}{2}\hbar\omega_{0}\coth\left(  \frac{\hbar\omega_{0}%
}{2k_{B}T}\right)  \right)  ^{2}-\left(  \frac{1}{2}\hbar\omega_{0}\right)
^{2}.
\end{align}
Once again the fluctuations in the energy of the harmonic oscillator differ
between classical and quantum theories. \ However, the discrepancy between
$\left\langle H^{2}\right\rangle _{T~classical.}$and $\left\langle \hat{H}%
^{2}\right\rangle _{T~quantum}$ in the classical and quantum theories at any
temperature $T$ involves exactly the same discrepancy which existed at zero
temperature,%
\begin{equation}
\left\langle H^{2}\right\rangle _{T~classical.}-\left\langle \hat{H}%
^{2}\right\rangle _{T~quantum}=\left(  \hbar\omega_{0}/2\right)  ^{2}.
\end{equation}
The positive temperature aspect brings the classical and the quantum
calculations closer together for the average value of the energy squared. \ On
the other hand, as the temperature goes to zero, we find $\left\langle \hat
{H}^{2}\right\rangle _{T~quantum}\rightarrow2\left(  \hbar\omega_{0}/2\right)
^{2}-\left(  \hbar\omega_{0}/2\right)  ^{2}=\left(  \hbar\omega_{0}/2\right)
^{2}=\left\langle \hat{H}\right\rangle _{0~quantum}^{2},$ which is the result
in Eq. (\ref{H2q}).

\section{Planck Spectrum of Quantum Thermal Radiation}

The change in the quantum mechanical procedure for the harmonic oscillator in
going from zero temperature to positive temperature is the same sort of shift
which appears for quantum thermal radiation. \ According to the textbooks of
modern physics and of quantum statistical mechanics,\cite{mod} the quantum
vacuum at zero temperature involves virtual radiation fluctuations but no real
photons. \ On the other hand, positive temperature indeed involves real
photons with a radiation energy per normal mode $\mathcal{E}_{P}(\omega
,T)\ $given by Planck's spectrum \textit{without} zero-point
radiation,\cite{mod}%
\begin{equation}
\mathcal{E}_{P}(\omega,T)=\frac{\hbar\omega}{\exp\left[  \hbar\omega
/(k_{B}T)\right]  -1}=\frac{1}{2}\hbar\omega_{0}\left[  \coth\left(
\frac{\hbar\omega_{0}}{2k_{B}T}\right)  -1\right]  . \label{EP}%
\end{equation}
Thus the quantum theory regards thermal radiation as vanishing at
zero-temperature, $\mathcal{E}_{P}(\omega,T)\rightarrow0$ as $T\rightarrow0$.
\ This is exactly the same view as was taken by the classical physicists
working with Boltzmann statistical mechanics at the end of the 19th century.
\ This viewpoint that the thermal spectrum vanishes at zero temperature is
quite different from the classical view which includes classical zero-point
radiation as an integral part of the spectrum of classical thermal
radiation.\cite{new} \ In classical theory, the random classical zero-point
radiation leads to the zero-point fluctuations of a classical
\textit{mechanical} oscillator which must match the average \textit{radiation}
energy at the natural frequency of the oscillator in order to be in
equilibrium with the random classical radiation. \ On the other hand, quantum
theory has an entirely different basis for thermal equilibrium between
radiation and matter. \ The quantum harmonic oscillator has a
\textit{mechanical} zero-point energy $\hbar\omega_{0}/2$ in Eq. (\ref{Eqq})
which does not appear in the quantum Planck \textit{radiation} spectrum Eq.
(\ref{EP}) which has no zero-point energy in quantum theory.

\section{Square of the Angular Momentum at Positive Temperature}

It is also interesting to see the extension of these ideas to the situation
involving the isotropic harmonic oscillator at strictly positive temperature
$T$.\cite{table} \ The classical theory treats all random radiation (both
zero-point radiation and thermal radiation above the zero-point radiation) as
thermal radiation contributing to the random particle motion on phase space,
as indicated in Eq. (\ref{PTxp}). \ At positive temperature, the classical
phase space remains spherically symmetric and corresponds to the identity
representation of the rotation group; accordingly, the classical Poisson
bracket of $L^{2}$ applied to the phase space remains $\left\{  P_{T}%
,L^{2}\right\}  =0,$ as in Eq. (\ref{L2P0}). \ The average value of the
angular momentum squared is still obtained by averaging over the phase space
as in Eq. (\ref{avclL2}) (but now with the phase space associated with
positive temperature as in Eq. (\ref{PTxp})) , and becomes
\begin{equation}
\left\langle L^{2}\right\rangle _{T~classical}=(3/2)\hbar^{2}\coth^{2}\left[
\hbar\omega_{0}/(2k_{B}T)\right]  .\label{L2cl}%
\end{equation}
\ 

On the other hand, for positive temperature $T>0$, the quantum theory no
longer represents the physical system in thermal equilibrium by a single
vector in Hilbert space, but rather treats the behavior associated with
strictly positive temperature in a fashion quite different from the
expectation value calculation in Eq. (\ref{avqL2}). \ Now a sum over the
excited eigenstates of energy $\mathcal{E}_{i}$ is required with a weighting
by the Boltzmann factor,
\begin{equation}
\left\langle \hat{L}^{2}\right\rangle _{T~quantum}=\sum\nolimits_{i}\left(
\int\nolimits_{-\infty}^{\infty}d^{3}x\psi_{i}^{\ast}\hat{L}^{2}\psi
_{i}\right)  \frac{\exp\left[  -\mathcal{E}_{i}/\left(  k_{B}T\right)
\right]  }{Z}.
\end{equation}
The average angular momentum operator squared involves the evaluation of terms
of the form\cite{table} $\left\langle \hat{x}^{2}\right\rangle _{T}=\left[
\hbar/(2m\omega_{0}\right]  \coth\left[  \hbar\omega_{0}/(2k_{B}T\right]
,~\left\langle \hat{p}^{2}\right\rangle _{T}=\left[  \hbar m\omega
_{0}/2\right]  \coth\left[  \hbar\omega_{0}/(2k_{B}T\right]  ,~\left\langle
\hat{x}\hat{p}\right\rangle _{T}=i\hbar/2,$ and $\left\langle \hat{p}\hat
{x}\right\rangle _{T}=-i\hbar/2\,.$ The result for the expectation value for
the square of the angular momentum is
\begin{equation}
\left\langle \hat{L}^{2}\right\rangle _{T~quantum}=(3/2)\hbar^{2}\coth
^{2}\left[  \hbar\omega_{0}/(2k_{B}T)\right]  -\left(  3/2\right)  \hbar^{2}.
\end{equation}
The quantum excited states no longer belong exclusively to the identity
representation of the rotation group with $l=0,$ but now include states with
representations with $l>0$, and so now give non-zero values for the angular
momentum squared. \ We notice that the difference between the classical and
the quantum average values of the angular momentum squared always differ by
the same amount $\left(  3/2\right)  \hbar^{2}$ which appeared at zero temperature%

\begin{equation}
\left\langle L^{2}\right\rangle _{T~classical}-\left\langle \hat{L}%
^{2}\right\rangle _{T~quantum}=\left(  3/2\right)  \hbar^{2}.
\end{equation}

At high temperature $\hbar\omega_{0}<<k_{B}T,$ we find the limit $\left(
\hbar\omega_{0}/2\right)  \coth\left[  \hbar\omega_{0}/(2k_{B}T\right]
\rightarrow k_{B}T\,,\ $so that the classical phase space distribution in Eq.
(\ref{PTxp}) becomes the Boltzmann probability distribution of traditional
classical statistical mechanics. \ The classical result for the average of the
square of the angular momentum in Eq. (\ref{L2cl}) goes over to $\left\langle
L^{2}\right\rangle _{T~classical}\approx6\left(  k_{B}T/\omega_{0}\right)
^{2}$ which is the value obtained from traditional classical statistical
mechanics where Planck's constant has completely disappeared from the
expression. \ The quantum result becomes $\left\langle \hat{L}^{2}%
\right\rangle _{T~quantum}\approx6\left(  k_{B}T/\omega_{0}\right)
^{2}-\left(  3/2\right)  \hbar^{2}$ which still subtracts the value $\left(
3/2\right)  \hbar^{2}$ associated with zero temperature. \ 

\section{Closing Summary}

In this article, we have explored some connections between classical and
quantum theories for the harmonic oscillator. \ As pointed out by Dirac, the
classical-quantum connection between Poisson brackets and quantum commutators
means that the theories agree on the basic algebraic structure associated with
groups of transformations. \ Dirac's connection to the quantum operators on
Hilbert space also gives well-defined numerical values tied to Planck's
constant $\hbar$ for the expectation values of the quantum operators. \ On the
other hand, the Poisson brackets provide canonical transformations, but no
numerical values for the physical quantities in the classical theory. \ The
classical theory must go outside nonrelativistic classical mechanics in order
to introduce a fundamental scale for the classical physical quantities. \ One
natural extension looks to classical electrodynamics, where Planck's constant
$\hbar$ can assume a natural role as the scale factor for the
Lorentz-invariant classical zero-point radiation which forms a natural part of
classical thermal radiation. \ The associated electromagnetic theory is often
termed \textquotedblleft stochastic electrodynamics,\textquotedblright\ and
can provide classical explanations for Casimir forces, van der Waals forces,
oscillator specific heats, diamagnetism, blackbody radiation, and the absence
of atomic collapse.\cite{SED} \ Any mechanical system located in random
classical radiation will inherit the radiation scale involving $\hbar$, and
will come to equilibrium with a phase space distribution corresponding to a
classical stochastic mechanical theory. \ We have used the classical
stochastic mechanical theory for our classical analysis in this article. \ For
the harmonic oscillator, the classical-quantum connections involve agreement
regarding average values, but disagreement regarding fluctuations. \ In all
cases, the classical behavior fits with fundamental classical ideas, which can
be quite different from quantum ideas. \ In some cases, the classical results
are reassuring. \ For example, the positive value for the square of the
angular momentum of the three-dimensional oscillator is more comforting to
classical sensibilities than is the zero-value claimed by the orthodox
interpreters of the quantum theory. \

\bigskip


\begin{thebibliography}{99}
\bibitem {table}We can calculate the results for $\left\langle L^{2}%
\right\rangle _{T}$ by using the average values for powers of displacements
and momenta at positive temperature quoted in Table II of ref. 21.
\end{thebibliography}
\end{document}